\documentclass[prb,twocolumn,showpacs,preprintnumbers,amsmath,aps]{revtex4}
\usepackage{graphicx,hyperref}
\usepackage{bm}
\usepackage{subfigure}

\def\nbC{{\mathchoice {\setbox0=\hbox{$\displaystyle\rm C$}%
\hbox{\hbox to0pt{\kern0.4\wd0\vrule height0.9\ht0\hss}\box0}}
{\setbox0=\hbox{$\textstyle\rm C$}\hbox{\hbox
to0pt{\kern0.4\wd0\vrule height0.9\ht0\hss}\box0}}
{\setbox0=\hbox{$\scriptstyle\rm C$}\hbox{\hbox
to0pt{\kern0.4\wd0\vrule height0.9\ht0\hss}\box0}}
{\setbox0=\hbox{$\scriptscriptstyle\rm C$}\hbox{\hbox
to0pt{\kern0.4\wd0\vrule height0.9\ht0\hss}\box0}}}}
\def\nbQ{{\mathchoice {\setbox0=\hbox{$\displaystyle\rm
Q$}\hbox{\raise
0.15\ht0\hbox to0pt{\kern0.4\wd0\vrule height0.8\ht0\hss}\box0}}
{\setbox0=\hbox{$\textstyle\rm Q$}\hbox{\raise
0.15\ht0\hbox to0pt{\kern0.4\wd0\vrule height0.8\ht0\hss}\box0}}
{\setbox0=\hbox{$\scriptstyle\rm Q$}\hbox{\raise
0.15\ht0\hbox to0pt{\kern0.4\wd0\vrule height0.7\ht0\hss}\box0}}
{\setbox0=\hbox{$\scriptscriptstyle\rm Q$}\hbox{\raise
0.15\ht0\hbox to0pt{\kern0.4\wd0\vrule height0.7\ht0\hss}\box0}}}}
\def\nbT{{\mathchoice {\setbox0=\hbox{$\displaystyle\rm
T$}\hbox{\hbox to0pt{\kern0.3\wd0\vrule height0.9\ht0\hss}\box0}}
{\setbox0=\hbox{$\textstyle\rm T$}\hbox{\hbox
to0pt{\kern0.3\wd0\vrule height0.9\ht0\hss}\box0}}
{\setbox0=\hbox{$\scriptstyle\rm T$}\hbox{\hbox
to0pt{\kern0.3\wd0\vrule height0.9\ht0\hss}\box0}}
{\setbox0=\hbox{$\scriptscriptstyle\rm T$}\hbox{\hbox
to0pt{\kern0.3\wd0\vrule height0.9\ht0\hss}\box0}}}}
\def\nbS{{\mathchoice
{\setbox0=\hbox{$\displaystyle     \rm S$}\hbox{\raise0.5\ht0%
\hbox to0pt{\kern0.35\wd0\vrule height0.45\ht0\hss}\hbox
to0pt{\kern0.55\wd0\vrule height0.5\ht0\hss}\box0}}
{\setbox0=\hbox{$\textstyle        \rm S$}\hbox{\raise0.5\ht0%
\hbox to0pt{\kern0.35\wd0\vrule height0.45\ht0\hss}\hbox
to0pt{\kern0.55\wd0\vrule height0.5\ht0\hss}\box0}}
{\setbox0=\hbox{$\scriptstyle      \rm S$}\hbox{\raise0.5\ht0%
\hboxto0pt{\kern0.35\wd0\vrule height0.45\ht0\hss}\raise0.05\ht0%
\hbox to0pt{\kern0.5\wd0\vrule height0.45\ht0\hss}\box0}}
{\setbox0=\hbox{$\scriptscriptstyle\rm S$}\hbox{\raise0.5\ht0%
\hboxto0pt{\kern0.4\wd0\vrule height0.45\ht0\hss}\raise0.05\ht0%
\hbox to0pt{\kern0.55\wd0\vrule height0.45\ht0\hss}\box0}}}}
\def\nbZ{{\mathchoice {\hbox{$\sf\textstyle Z\kern-0.4em Z$}}
{\hbox{$\sf\textstyle Z\kern-0.4em Z$}}
{\hbox{$\sf\scriptstyle Z\kern-0.3em Z$}}
{\hbox{$\sf\scriptscriptstyle Z\kern-0.2em Z$}}}}

\begin{document}

\title{Critical scaling in random-field systems: 2 or 3 independent exponents?}

\author{Gilles Tarjus} \email{tarjus@lptmc.jussieu.fr}
\affiliation{LPTMC, CNRS-UMR 7600, Universit\'e Pierre et Marie Curie,
bo\^ite 121, 4 Pl. Jussieu, 75252 Paris c\'edex 05, France}

\author{Ivan Balog} \email{balog@ifs.hr}
\affiliation{LPTMC, CNRS-UMR 7600, Universit\'e Pierre et Marie Curie,
bo\^ite 121, 4 Pl. Jussieu, 75252 Paris c\'edex 05, France}
\affiliation{Institute of Physics, P.O.Box 304, Bijeni\v{c}ka cesta 46, HR-10001 Zagreb, Croatia}

\author{Matthieu Tissier} \email{tissier@lptmc.jussieu.fr}
\affiliation{LPTMC, CNRS-UMR 7600, Universit\'e Pierre et Marie Curie,
bo\^ite 121, 4 Pl. Jussieu, 75252 Paris c\'edex 05, France}

\date{\today}

\begin{abstract}
We show that the critical scaling behavior of random-field systems with short-range interactions and disorder correlations 
cannot be described in general by only two independent exponents, contrary to previous claims. This conclusion is based 
on a theoretical description of the whole $(d,N)$ domain of the $d$-dimensional random-field $O(N)$ model and points to 
the role of rare events that are overlooked by the proposed derivations of two-exponent scaling. Quite strikingly, however, 
the numerical estimates of the critical exponents of the random field Ising model are extremely close to the predictions of 
the two-exponent scaling, so that the issue cannot be decided on the basis of numerical simulations.

\end{abstract}

\pacs{11.10.Hi, 75.40.Cx}

\maketitle

\section{Introduction}

The critical behavior of models in the presence of a quenched random field has attracted a lot of attention over the past decades.
\cite{imry-ma75,nattermann98} Beyond the experimental interest, such models provide a rich playground to investigate the influence of 
quenched disorder on the long-distance properties of a system. Their equilibrium critical point, separating a disordered, paramagnetic 
phase from an ordered, ferromagnetic one, is known to be controlled by a zero-temperature fixed point, with temperature being an 
irrelevant variable in the renormalization-group sense. Among the new features brought by such a fixed point is the violation of the 
``hyperscaling relation'' between critical exponents. The space dimension $d$ that appears in this relation must be corrected by the  
exponent $\theta$ describing the flow of the renormalized temperature to zero, thereby leading to an apparent dimensional reduction 
from $d$ to $d-\theta$.\cite{nattermann98,villain85,fisher86}

Whereas phenomenological theories take $\theta$ as an independent exponent,\cite{villain85,fisher86} which implies that equilibrium 
scaling behavior is described by three independent exponents in place of the usual two-exponent scaling for finite-temperature fixed points, 
two different approaches have claimed that $\theta$ is actually fixed. One is the dimensional-reduction prediction, based either on 
perturbation theory\cite{aharony76,grinstein76,young77} or on the Parisi-Sourlas supersymmetric formulation.\cite{parisi79} 
It states that $\theta=2$ and that the exponents of the random-field model are those of the corresponding pure model in two dimension less. 
This prediction has been rigorously proven to be wrong in low enough dimension,\cite{imbrie84,bricmont87} and we have recently 
provided a complete theoretical explanation of dimensional reduction and its breakdown through a nonperturbative functional 
renormalization group.\cite{tarjus04,tissier06,tissier11,tarjus13}

Another line of argument has been put forward by Schwartz and coworkers.\cite{schwartz85,schwartz86} The claim is that $\theta=2-\eta$, 
with $\eta$ the anomalous dimension of the (order parameter) field, so that scaling is described by only two independent exponents, \textit{e.g.}, $\eta$ and 
the correlation length exponent $\nu$. The derivation involves general manipulations and the result is supposed to hold for the 
Ising as well as the continuous version with $O(N)$ symmetry. It is also supported by heuristic considerations.\cite{schwartz86,binder}

The problem with the above derivations is that they rely on formal manipulations that are blind to the presence of rare events or rare regions, 
such as avalanches or droplets, or that overlook the presence of multiple metastable states. This is known to be the reason for the failure of 
the simple supersymmetric formulation.\cite{parisi84,tissier11,tarjus13} This also casts some doubt on the general validity of 
the order-of-magnitude estimates of the relative strength of the fluctuations carried out in Schwartz's arguments.\cite{schwartz85,schwartz86}

How to conclude on the validity of the two-exponent scenario proposed by Schwartz and coworkers? The question is actually quite subtle as 
the description is exact at the lower and the higher critical dimensions of the random-field models and it appears to be numerically very well verified 
in computer simulations (including a recent extensive one\cite{fytas13}), 
high-temperature expansions or approximate renormalization-group treatments, mostly in $d=3$. Our contention is 
that it is impossible to answer the question on a pure numerical ground as the error bars will always blur the conclusion. On the contrary, with 
the help of the functional renormalization group (FRG), the problem can be studied for continuous values of the dimension $d$ and of the number of 
components $N$ of the order-parameter field. We can then show with no ambiguity that the two-exponent scenario cannot be right in general, which then reduces 
the corresponding prediction for the random-field Ising model in $d=3$ to a very, or even extremely,\cite{fytas13} good but by no 
means exact, result. We have already made this point in earlier work\cite{tissier06,tissier11} but in this paper, we sharpen the arguments 
and present extended calculations.

The rest of the paper is organized as follows. In section II, we present the the random field $O(N)$ model [RF$O(N)$M] and the scaling 
behavior around its critical point and in section III we summarize and discuss the predictions of the two-exponent scaling scenario. 
In section IV, we recall the results of the FRG for the RF$O(N)$M with $N>1$ (continuous symmetry) near the lower critical dimension for ferromagnetism, 
in $d=4+\epsilon$. We show that at order $\epsilon$ and $\epsilon^2$, the relation proposed by Schwartz and coworkers is unambiguously 
violated. In the next section, we consider the random field Ising model (RFIM) via a nonperturbative version of the FRG that 
we have previously developed. Here too, we show that the two-exponent scenario cannot be right for all dimensions between $2$ and $6$. 
We focus in particular on the dimensions around the critical dimension $d_{DR}\simeq 5.1$ that marks the lower limit of existence of the 
supersymmetric fixed point associated with the $d \rightarrow d-2$ dimensional reduction. Finally, we give 
some concluding remarks in section V.

\section{The critical behavior of random field systems}
\label{scaling}

The long-distance behavior of the random field $O(N)$ model is described by the following Hamiltonian or bare action,
\begin{equation}
\label{eq_ham_rfon}
S[\bm{\varphi},\mathbf h]= \int_{x} \bigg\{ \frac{1}{2}  \vert\partial \bm\varphi(x) \vert ^2 + U_B(\vert \bm\varphi(x)\vert^2) -
\mathbf h(x)\bm{.} \bm{\varphi}(x) \bigg\} ,
\end{equation}
where $ \int_{x} \equiv \int d^d x$,  $\bm\varphi(x)$ is an $N$ component field, $U_B( \vert\bm\varphi \vert ^2)= (\tau/2)  
\vert \bm\varphi \vert ^2  +  (u/4!) ( \vert \bm\varphi\vert ^2)^2$, 
and $\mathbf h(x)$ is a random source (a random magnetic field in the language of magnetic systems) with zero mean and a variance
\begin{equation}
\label{eq_cum_dis}
\overline{h^{\mu}(x)h^{\nu}(y)}=\Delta_B \ \delta_{\mu\nu}\delta^{(d)}(x-y)
\end{equation}
where $\mu,\nu=1, \cdots,N$ and an overline denotes an average over the random field. An ultra-violet (UV) momentum cutoff $\Lambda$, 
associated with an inverse microscopic length scale such as a lattice spacing, is also implicitly considered. We focus here on the 
\textit{short-range} version defined above. We shall briefly comment on the long-range version in the conclusion.

Due to the presence of the random field, one needs to consider two different types of pair correlation functions of the $\bm\varphi$ field: 
the so-called connected one, $G_{con}^{\mu\nu}(x-y)=\overline{\langle \varphi^{\mu}(x)\varphi^{\nu}(y)\rangle-\langle \varphi^{\mu}(x)\rangle
\langle\varphi^{\nu}(y)\rangle}$ and the disconnected one, $G_{dis}^{\mu\nu}(x-y)=\overline{\langle \varphi^{\mu}(x)\rangle\langle 
\varphi^{\nu}(y)\rangle}-\overline{\langle \varphi^{\mu}(x)\rangle}\; \overline{\langle\varphi^{\nu}(y)\rangle}$. At the critical point $T_c$, the two correlation functions behave as
\begin{equation}
\begin{aligned}
\label{eq_con_disc_critical}
&G_{con}^{\mu\nu}(x-y)\sim  \frac{T\,\delta_{\mu\nu}}{\vert x-y\vert^{d-2+\eta}} \;,\\&
G_{dis}^{\mu\nu}(x-y)\sim  \frac{\delta_{\mu\nu}}{\vert x-y\vert^{d-4+\bar\eta}} \;,
\end{aligned}
\end{equation}
where $\eta$ is the usual anomalous dimension of the field and $\bar\eta$ is \textit{a priori} a new exponent. Accordingly, one can also 
define two types of susceptibilities that diverge as one approaches the critical point from above as
\begin{equation}
\begin{aligned}
\label{eq_con_disc_susceptibilities}
&\chi_{con}=\int_xG_{con}(x) \sim \left (T-T_c\right )^{-\gamma} \;,\\&
\chi_{dis}=\int_xG_{dis}(x) \sim \left ( T-T_c\right )^{-\bar\gamma} \;,
\end{aligned}
\end{equation}
the former one being the usual magnetic susceptibility. [The component indices have been dropped as the functions are then proportional 
to $\delta_{\mu\nu}$ as in Eq. (\ref{eq_con_disc_critical}).] The exponents $\gamma$ and $\bar\gamma$ are related to $\eta$ and $\bar\eta$ 
via $\gamma=(2-\eta)\nu$ and $\bar\gamma=(4-\bar\eta)\nu$, with $\nu$ the correlation length exponent.

The renormalized temperature is irrelevant at the fixed point controlling the critical behavior\cite{villain85,fisher86} and it 
flows to zero with an exponent $\theta$. As a result, the hyperscaling relation has an unusual form,
\begin{equation}
\label{eq_hyperscaling}
2-\alpha=(d-\theta)\nu \;,
\end{equation}
where $\alpha$ is the specific-heat  exponent. The exponents $\theta$, $\eta$, and 
$\bar\eta$ are related through
\begin{equation}
\label{eq_theta_etabar}
\theta=2+\eta-\bar\eta \;,
\end{equation}
so that the scaling around the critical point is \textit{a priori} described by three independent exponents, \textit{e.g.}, $\eta$, $\nu$, and 
$\theta$ or $\bar\eta$.

\section{The two-exponent scaling description}
\label{two-exponent}

The $d\rightarrow d-2$ dimensional reduction predicts that $\theta=2$, \textit{i.e.} $\bar\eta=\eta$, and furthermore that all the 
critical exponents are given by those of the pure model in $(d-2)$. On the other hand, the two-exponent scenario put forward by 
Schwartz and coworkers\cite{schwartz85,schwartz86} states that the exponents obey the following relations:
\begin{equation}
\label{eq_two_exponent}
\theta=2-\eta\, , \;\;\; \bar\eta=2\eta\,.
\end{equation}
The derivation actually also implies that the disconnected and connected correlation functions in Fourier space are related through
\begin{equation}
\begin{aligned}
\label{eq_disc_con}
G_{dis}(q)=\frac{\Delta_B}{T^2}\,G_{con}(q)^2
\end{aligned}
\end{equation}
at criticality and when $q\rightarrow 0$. Eq.~(\ref{eq_disc_con}) implies that the second cumulant of the random field is not renormalized 
and stays fixed to its bare value $\Delta_B$.

A stronger claim was initially made by Schwartz,\cite{schwartz85} who suggested that all the 
exponents of a random-field system in dimension $d$ are the same as those of its pure counterpart in a reduced dimension 
$d-2+\eta(d)$. This prediction was however soon shown by Bray and Moore\cite{bray85} to be already wrong for the exponent $\nu$ 
in the RFIM near its lower critical dimension 2 at first order in $\epsilon=d-2$. To the best of our knowledge, it was subsequently abandoned.

It is also worth  mentioning that various heuristic arguments have been used to derive the above two-exponent scaling.\cite{schwartz86,binder} 
For instance, one such argument states that the magnetization per spin in a finite-size system of linear size $L$ (we consider here the RFIM 
for simplicity), which scales as $L^{-(d-4+\bar\eta)/2}$, is given for a typical realization of the random field by the average magnetic 
susceptibility $\chi_{con,L}\sim L^{2-\eta}$ times the mean random field $\overline h_L$ which scales as $L^{-d/2}$: 
\begin{equation}
\label{eq_binder}
m_L\sim L^{-(d-4+\bar\eta)/2}\sim \chi_{con,L}\, \overline h_L\sim L^{2-\eta}L^{-d/2}\,.
\end{equation}
It immediately results that $\bar\eta=2\eta$. 

In the case of the RFIM, the two-exponent scenario is exact at and near the lower critical dimension in first order in $\epsilon=d-2$, 
as shown in Ref. [\onlinecite{bray85}], and it is also somewhat trivially expected at the upper dimension $d=6$ and in first 
order in $\epsilon=6-d$ as both $\eta$ and $\bar\eta$ are zero and the mean-field result $\theta=2$ still  applies. In between, numerical studies 
via high-temperature expansions in $d=3,4,5$\cite{gofman93} and computer simulations in $d=3$, confirm that $\bar\gamma$ is very close 
to $2\gamma$, and $\bar\eta$ to $2\eta$, certainly within the accuracy of the methods.

However, as mentioned in the Introduction, a fundamental problem remains that the proposed derivations of the two-exponent scaling involve, beyond 
formal manipulations, estimates of the relative order of magnitude of the fluctuations that essentially rely on factorization approximations and 
the central-limit theorem in the limit of large system size (see \textit{e.g.} the above heuristic argument). All of the derivations 
are therefore blind to rare events, rare regions or rare samples, which, precisely, have been shown to be crucial in disordered systems near zero-temperature fixed points.\cite{villain85,fisher86,droplet_bray,droplet_fisher,fisherFRG,balents96,FRGledoussal-chauve,BLbalents05,
FRGledoussal-wiese,static-distrib_ledoussal,BLledoussal10,tissier06,tissier11,tarjus13}. 

In the absence of rigorous derivations, numerical evidence, no matter how good, is insufficient to establish 
the validity of the scenario, because of unavoidable uncertainty. Quite the contrary, we show below that the two-exponent description, which is 
claimed to apply to random-field systems below their upper critical  dimension irrespective of the number of components $N$ and 
of the dimension $d$, cannot be valid in general and, as a consequence, has no rigorous foundations at this point.

\section{The RFO(N)M near $d=4$}
\label{sec:RFO(N)M}

Near the lower critical dimension for ferromagnetism, which is equal to $d=4$ in this case, the long-distance physics of the RF$O(N)$M is captured 
by a nonlinear $\sigma$ model that in turn can be studied through a perturbative but functional RG.\cite{fisher85} The resulting FRG flow equations in 
$d=4+\epsilon$ have been studied at one-\cite{fisher85,feldman01,tissier06} and two-\cite{tissier06b,ledoussal06} loop order.

The dimensionless second cumulant of the renormalized random field is a function $\Delta(z)$ (where $z$ is the cosine of the 
angle between fields in two different copies of the system\cite{fisher85,tissier06}) and it obeys the following RG equation at one loop:
\begin{equation}
  \label{eq_Delta(z)}
\begin{aligned}
&k\partial_k \Delta_k(z) = \epsilon  \Delta_k(z) - \bigg[z \Delta_k(z)^2+(N-3) \Delta_k(1) \Delta_k(z)\\&
+ (N-3+4z^2)\Delta_k(z)\Delta_k'(z) -(N+1)z\Delta_k(1)\Delta_k'(z)\\&
-z(1-z^2)\Delta_k(z)\Delta_k''(z)+(1-z^2)\Delta_k(1)\Delta_k''(z)\\&
-3z(1-z^2)\Delta_k'(z)^2+(1-z^2)^2\Delta_k'(z)\Delta_k''(z)\bigg]\,,
 \end{aligned}
\end{equation}
where $k$ is the running infrared (IR) momentum cutoff, a prime denotes a derivative, and $\Delta_k(z)$ 
[which, up to a multiplicative constant, was denoted $R'(z)$ in previous publications] 
is of order $\epsilon$ near the fixed point. One can moreover define two running exponents $\eta_k$ and $ \bar \eta_k$ as follows:
\begin{equation}
  \label{eq_eta_fisherfinal}
\eta_k = \Delta_k(1), \; \bar \eta_k = -\epsilon + (N-1) \Delta_k(1)\,.
\end{equation}
They converge to the fixed point values $\eta$, $\bar\eta$ when $k\rightarrow 0$. The corresponding equations at 
two-loop order are given in Ref. [\onlinecite{tissier06b}] and are not reproduced here.

A numerical and analytical investigation of these FRG equations\cite{tissier06,tissier06b,baczyk13b} shows that above a critical value 
of the number of components, $N_{DR}=18-\frac{49}{5}\epsilon$, there exists a fixed point that corresponds to the $d\rightarrow d-2$ 
dimensional reduction, with $\bar\eta=\eta= \frac{\epsilon}{N-2}-\frac{(N-1)\epsilon^2}{(N-2)^2}$. Above a slightly higher value, 
$N_{cusp}=2(4+3\sqrt 3)-3(2+3\sqrt{3})\epsilon/2$,\cite{baczyk13b} this fixed point is stable and 
describes the critical behavior of the RF$O(N)$M (see also Ref. [\onlinecite{sakamoto06}]). In this domain of $N$, critical scaling is 
therefore described by $\bar\eta=\eta$ and $\theta=2$, which contradicts the predictions of Schwartz and coworkers.\cite{schwartz85,schwartz86}

Below $N_{cusp}$, the stable fixed point is now characterized by a nonanalyticity in the functional dependence of the 
renormalized disorder cumulant [$\Delta_*(z)-\Delta_*(1)\sim \sqrt{1-z}$ when $z\rightarrow 1$] that is strong enough to break the dimensional-reduction 
prediction, with therefore $\bar\eta>\eta$. However, the possibility that this nonanalytic ``cuspy'' fixed point is characterized by 
the equality $\bar\eta=2\eta$ is invalidated by the results. We display the ratio $(2\eta-\bar\eta)/\eta$ as a function of $N$ at one-loop order 
in Fig. 1. (The results are confirmed at two-loop order.) The ratio is equal to 1 above $N_{cusp}\simeq 18.3923$ and decreases continuously 
 as $N$ decreases. At the endpoint value at which $\Delta_*(1)$ diverges, $N_c=2.8347\cdots$, 
it reaches a strictly positive value of $3-N_c\simeq 0.1653\cdots$. We stress that the output can be obtained with an arbitrary precision, 
which is quite different from simulations.

\begin{figure}[ht]
\includegraphics[width=\linewidth]{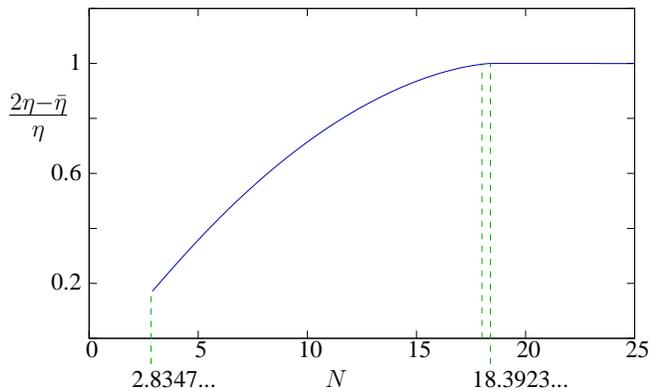}
\caption{ Ratio $(2\eta-\bar\eta)/\eta$ as a function of $N$ for the RF$O(N)$M in $d=4+\epsilon$ at one-loop order. Above 
$N_{cusp}=2(4+3\sqrt 3)\simeq 18.3923\cdots$ the stable fixed point is associated with the $d\rightarrow d-2$  dimensional reduction 
and $(2\eta-\bar\eta)/\eta=1$. Below $N_{DR}=18$ (marked by the leftmost vertical dashed line), the critical behavior is 
described by a ``cuspy'' fixed point and dimensional reduction 
breaks down but $(2\eta-\bar\eta)/\eta$ is generically $>0$. In a small interval $N_{DR}<N<N_{cusp}$ the two fixed points coexist, but 
the dimensional-reduction one is unstable to a nonanalytic, ``cuspy'' perturbation.}
\end{figure}

\section{The RFIM near $d_{DR}\simeq 5.1$}
\label{sec:RFIM}

The main result proving that the two-exponent scaling is generically inexact in the short-range RFIM is that there exists a range 
of dimension below the upper critical one, $d=6$, for which the $d\rightarrow d-2$ dimensional reduction is valid and 
therefore for which $\bar\eta=\eta \neq 2\eta$ and $\theta=2 \neq 2-\eta$. The arguments establishing this result are as follows:

(i) The fixed point associated with dimensional reduction proceeds \textit{continuously} from the Gaussian fixed point in $d=6$ as one lowers $d$. 
It is characterized by the absence of a linear cusp in the functional dependence of the cumulants of the renormalized random field. (It can 
indeed be shown by the FRG that only such a cusp can lead to a breakdown of dimensional reduction.\cite{tarjus04,tissier11}) 

(ii) That this  dimensional-reduction fixed point is stable for some interval of $d$ below $6$ can furthermore be seen by studying, 
in addition to all possible ``cuspless'' perturbations that indeed prove to be irrelevant, the 
eigenvalue associated with a nonanalytic, ``cuspy''  perturbation around this fixed point. This can be done exactly  
in $d=6$ and can be numerically obtained below $6$. The perturbation is then found to be irrelevant at and in the vicinity of 
$d=6$: the eigenvalue is equal to $1$ in $d=6$ and \textit{continuously} decreases as one lower $d$ slightly below 6.\cite{tarjus13}

(iii) By means of a nonperturbative truncation of the exact hierarchy of FRG equations (NP-FRG) for the cumulants of the renormalized 
disorder, we have located the limit of existence of the dimensional-reduction fixed point at $d_{DR}\simeq 5.1$.\cite{tissier11} 
There is thus a nonzero range of dimension above $d_{DR}$ and below 6 in which $\bar\eta=\eta\neq 0$ and $\theta=2$, 
which contradicts the claim of the two-exponent scaling scenario.

In addition, we have investigated in more detail the ``cuspy''  fixed point corresponding to a breakdown of the $d\rightarrow d-2$ 
dimensional reduction below $d_{DR}$. Our purpose was to show that this fixed point is  
described by three independent exponents in general. The NP-FRG equations that must be numerically solved are given in 
Ref. [\onlinecite{tissier11}], together with technical comments, and this is not reproduced here.

We focus on the vicinity of $d_{DR}$, which is where the violation of the proposed relation $\bar\eta=2\eta$ is unambiguous. 
We display in Fig. 2 for illustration the RG flow of the running exponents $\eta_k$ and $\bar\eta_k$ (for a crisper visualization, 
we actually plot the combination $2\eta_k-\bar\eta_k$) in $d=5$ as a function of $\log(k/\Lambda)$, where $\Lambda$ 
is the UV cutoff and $k$ the running IR scale. When starting from initial conditions at the UV scale that are 
analytic (as the bare action), $\eta_k$ and $\bar\eta_k$ first equal each other until one reaches a scale, the so-called ``Larkin scale'',
\cite{FRGledoussal-chauve,FRGledoussal-wiese,tarjus04}
at which a strong enough nonanalyticity appears in the functional dependence of the second cumulant of the renormalized random 
field and the two exponents $\eta_k$ and $\bar\eta_k$ start to deviate. However, as can be seen, $d=5$ is close to $d_{DR}$, so  
that the difference between the exponents remains small at the fixed point ($k\rightarrow 0$) and $2\eta -\bar\eta$ is strictly larger 
than 0.

\begin{figure}[ht]
\includegraphics[width=\linewidth]{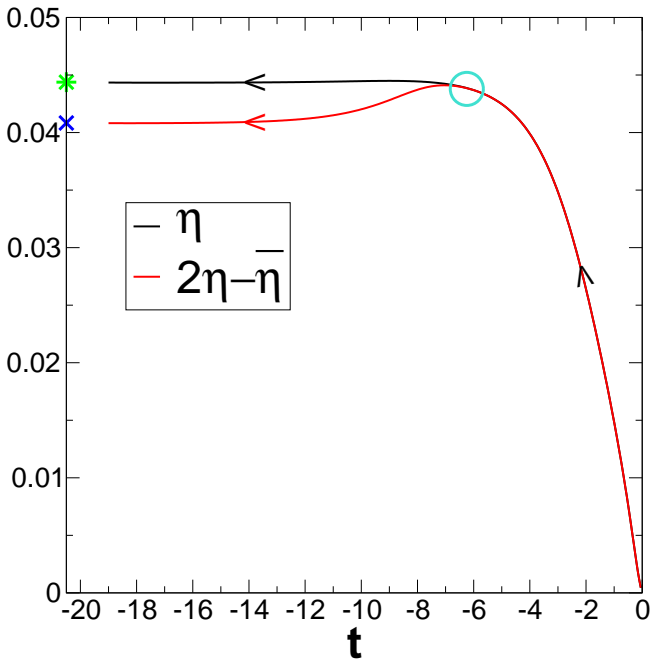}
\caption{ NP-FRG flow of the running exponent $\eta_k$ and of $2\eta_k-\bar\eta_k$ as a function of $t=\log(k/\Lambda)$ 
for the RFIM in $d=5$. The arrows indicate the flow toward the infrared ($k\rightarrow 0$). Note that one reaches $t\sim -19$, 
which means a length scale bigger than $10^8$ times the microscopic one. One is 
then clearly in the asymptotic regime as shown by comparing the curves with the symbols on the $y$-axis, which  are the values 
obtained by directly solving the fixed-point equation. The two curves are superimposed, implying $\bar\eta_k=\eta_k$, until the 
Larkin scale (marked by a circle), at which point they separate. However, the fixed point values are such that 
$2\eta-\bar\eta$ is unambiguously strictly greater than zero.}
\end{figure}

We also plot in Fig. 3 the fixed-point value of the ratio $(2\eta-\bar\eta)/\eta$ as a function of $d$ for the RFIM in the vicinity of $d_{DR}$ 
from the NP-FRG equations.\cite{footnote0} The crux of the present demonstration is not the actual values of the exponents (which, however, are 
always within 10$\%$ of the best estimates), but the fact that, around $d_{DR}$, $\bar\eta$ and $\eta$ are equal or close to each other 
so that $(2\eta-\bar\eta)/\eta$ stays near 1 and unambiguously violates the prediction of the two-exponent scenario. This is clearly seen in Fig. 3.

This conclusion is similar to that reached above for the RF$O(N)$M with continuous symmetry near $d=4$ and it may actually 
be extended through the NP-FRG to the whole $(N,d)$ domain\cite{tissier06} (note that the approximate nonperturbative 
truncation of the exact FRG equations gives back the exact perturbative FRG treatment in $d=4+\epsilon$ at one loop).

\begin{figure}[ht]
\includegraphics[width=\linewidth]{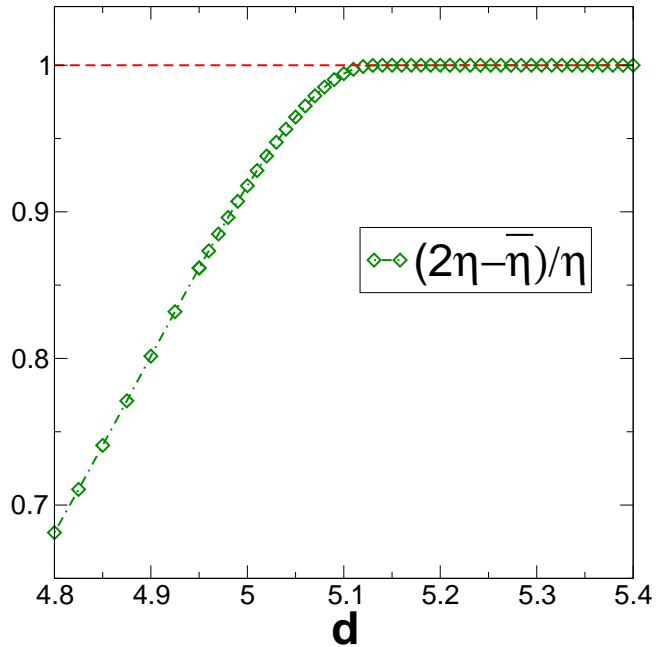}
\caption{ Ratio $(2\eta-\bar\eta)/\eta$ as a function of $d$ for the RFIM in the vicinity of $d_{DR}\simeq 5.1$ from the NP-FRG. 
Above $d_{DR}$, the ratio is exactly equal to 1 as the stable fixed point is associated with the $d\rightarrow d-2$  dimensional reduction. 
For $d<d_{DR}$, the stable fixed point is characterized by the existence of a cusp in the cumulants of the renormalized random field. The 
ratio then decreases continuously from 1 as one decreases $d$. As in Fig. 1, there is a finite domain of $d$ for which  
$2\eta-\bar\eta \neq 0$ with no ambiguity.}
\end{figure}

\section{Conclusion}

In this paper, we have challenged the two-exponent scaling scenario of the critical behavior of random-field systems proposed by 
Schwartz and coworkers.\cite{schwartz85,schwartz86} Due to the generality of its derivations, this scenario is supposed to apply to 
all random field models, whether in the Ising or in the $O(N)$ version, and in all dimensions $d$ between the lower and the upper critical ones. 
We have however clearly proven that the predictions, most notably the relation between exponents $\bar\eta=2\eta$, cannot be  
exact for generic random-field systems.\cite{footnote1} Our proof is beyond error bars as it involves the fact that there is a whole domain of $d$ 
and $N$, including the Ising version, for which instead $\bar\eta=\eta\neq 0$ and that near the boundary of this domain the exponents verify 
$\bar\eta\simeq \eta\neq 2\eta$.

We have pointed out that a potential caveat of the derivations of the two-exponent scaling, which imply that the random-field strength is not 
renormalized by fluctuations at criticality, is that they are blind to rare events, regions or samples that are known to play an important 
role in random-field systems. The arguments leading to two-exponent scaling however apply to random-field 
systems with long-range correlated disorder. In such systems, the bare variance of the random field is no longer given by Eq.~(\ref{eq_cum_dis}) 
but is characterized by an exponent $\rho$:
\begin{equation}
\label{eq_cum_disLR}
\overline{h^{\mu}(x)h^{\nu}(y)}\sim \Delta_B \frac{\delta_{\mu\nu}}{\vert x-y \vert^{d-\rho}}\,.
\end{equation}
The long-range piece of the second cumulant of the random field is then {\it not} renormalized by fluctuations.\cite{bray86b,fedorenko07,baczyk13} 
In this case, one therefore rigorously has that $2\eta-\bar\eta=\rho$. This is for instance correctly given by the heuristic derivation leading 
to Eq. (\ref{eq_binder}) as the mean random field in a finite system scales as $L^{-(d-\rho)/2}$. Similarly, a generalization of Eq. (\ref{eq_disc_con}) 
is now also valid. 

 In the case of the short-range RFIM, the two-exponent scaling description is exact close to the lower and 
(more trivially) upper critical dimensions. It moreover appears to provide an extremely good approximation in between, 
with $2\eta-\bar\eta \sim 10^{-2}$ or less. In the light of our results, this implies that computer simulations, even elaborate 
and extensive ones as the recent study in [\onlinecite{fytas13}], 
will likely be always inconclusive due to unavoidable residual errors.\cite{footnote2}


\begin{thebibliography}{15}
\bibitem{footnote0}
Note that in the case of the RFIM, $d_{DR}$ and $d_{cusp}$ appear essentially indistinguishable.\cite{baczyk13b} The difference could 
only affect a very small region near $d\simeq 5.1$ with no consequences on the present demonstration.
\bibitem{footnote1}
On the other hand, the Schwartz-Soffer inequality, $2\eta-\bar\eta \geq0$,\cite{schwartz-soffer85} which has a rigorous basis, 
is of course not questioned here.
\bibitem{footnote2}
For instance, a recent extensive numerical study of the critical behavior of the 3-$d$ RFIM via ground-state determination at $T=0$,\cite{fytas13} 
states that $2\eta-\bar\eta \sim 2-3\times 10^{-3}$. Our contention is that this does not prove, nor actually disprove, that $\bar\eta=2\eta$ 
despite the very small value.
\end{thebibliography}
\end{document}